\documentclass[aps,prc,twocolumn,showpacs,floatfix,nofootinbib,preprintnumbers,superscriptaddress,amsmath,amssymb,longbibliography,amsproc]{revtex4-2}
\usepackage[utf8]{inputenc}
\usepackage[T1]{fontenc}

\usepackage{graphicx}
\usepackage{amsmath}
\usepackage{amssymb}
\usepackage{dcolumn}
\usepackage{multirow}
\usepackage[colorlinks=true,linkcolor=blue,citecolor=blue,urlcolor=blue]{hyperref}
\usepackage{bbold}
\usepackage{bibentry}
\usepackage{orcidlink}
\usepackage{microtype}
\usepackage{mathtools,slashed}

\newcommand{\ket}[1]{\left| #1 \right\rangle}
\newcommand{\braket}[2]{\left\langle #1 \vert #2 \right\rangle}

\newcommand{\nopt}{\ensuremath{N_{\rm opt}}}
\newcommand{\Nshots}{\ensuremath{N_{\rm shots}}}
\newcommand{\Nexp}{\ensuremath{N_{\rm e}}}

\begin{document}

\title{Toward scalable quantum computations of atomic nuclei}

\author{Chenyi Gu\orcidlink{0000-0002-9346-3949}}
\affiliation{Department of Physics and Astronomy, University of
  Tennessee, Knoxville, Tennessee 37996, USA}

\author{Matthias Heinz\orcidlink{0000-0002-6363-0056}}
\email[Corresponding author: ]{heinzmc@ornl.gov}
\affiliation{National Center for Computational Sciences, Oak Ridge National Laboratory, Oak Ridge, Tennessee 37831, USA}
\affiliation{Physics Division, Oak Ridge National Laboratory, Oak
  Ridge, Tennessee 37831, USA}

\author{Oriel~Kiss\orcidlink{0000-0001-7461-3342}}
\affiliation{European Organization for Nuclear Research (CERN), Geneva 1211, Switzerland}
\affiliation{Department of Nuclear and Particle Physics, University of Geneva, Geneva 1211, Switzerland} 

\author{Thomas~Papenbrock\orcidlink{0000-0001-8733-2849}}
\affiliation{Department of Physics and Astronomy, University of Tennessee, Knoxville, Tennessee 37996, USA}
\affiliation{Physics Division, Oak Ridge National Laboratory, Oak Ridge, Tennessee 37831, USA}

\begin{abstract}
We solve the nuclear two-body and three-body bound states via quantum simulations of pionless effective field theory on a lattice in position space. While the employed lattice remains small, the usage of local Hamiltonians including two- and three-body forces ensures that the number of Pauli terms scales linearly with increasing numbers of lattice sites. We use an adaptive ansatz grown from unitary coupled cluster theory to parametrize the ground states of the deuteron and $^3$He, compute their corresponding energies, and analyze the scaling of the required computational resources.
Our quantum simulations reproduce exact benchmarks for $^2$H and $^3$He within 100~keV, requiring at most 30 layers in the ansatz and thus resulting in modest circuit depths.
Additionally, we find the number of shots required to reach a given precision scales linearly in the lattice size and more mildly in the system size.
Based on the agreement with exact benchmarks and mild scaling, we conclude that this can be an efficient, scalable approach for quantum computations of nuclear ground states, particularly to prepare initial states for quantum phase estimation or other filtering algorithms.
\end{abstract}
\maketitle

\section{Introduction}

In recent years, quantum computing has attracted considerable interest in nuclear theory. Given the limited coherence times of current quantum devices, research has focused on solving nuclear  models~\cite{peruzzo2014,dumitrescu2018,klco2018,cervia2021,dimatteo2021,stetcu2022,kiss2022,hlatshwayo2022,balantekin2023,perezobiol2023,robin2023,Grossi_Lipkin,singh2025advancing,Schoulepnikoff_EF_shell,Bhoy:2024uuh,Siwach_nuclear_gray_encoding,Sarma_23_oxygen,yoshida2025bridging,chinnarasu2025}, developing computational methods and algorithms~\cite{roggero2018,Lee:2019zze,lacroix2020,holland2020,choi2021,Lacroix:2022vmg,stetcu2023}, tackling classically ``hard'' problems like nuclear scattering and dynamics~\cite{kharzeev2020,turro2022,davoudi2023,bedaque2022,sharma2024,hall2021,watkins2024,wang2024,rethinasamy2024,weiss2025}, and gaining deeper insights into entanglement~\cite{beane2019,robin2021,gu2023}. For recent reviews, we refer the reader to Refs.~\cite{klco2022,review_Ramos,review_savage}.

While this body of work reflects the excitement surrounding a new technology, more sober assessments~\cite{lee2023,hoefler2023} raise doubts about whether a practical quantum advantage can be realized in computing ground-state energies. The skepticism stems from two main points. First, it is in general difficult to prepare initial states with a large overlap with the ground state. Second, many classical algorithms scale polynomially with system size and are sufficiently accurate to compute various properties of nuclear ground states~\cite{yao2020,stroberg2021,hu2022,elhatisari2024,sun2025,door2025}. Such computations have advanced tremendously in scale and sophistication—beyond what was deemed possible just a decade ago.

Nevertheless, quantum algorithms offer a stringent theoretical advantage: They can provide guaranteed solutions within a specified error tolerance, something no classical method can ensure. This is typically achieved via quantum phase estimation (QPE)~\cite{kitaev1995} and its early fault-tolerant variants~\cite{LinTong_2022,LinTong_qetu,Stat_QPE,Kiss2025earlyfaulttolerant}. The cost of QPE is determined by two main components: the preparation of an approximate initial state~\cite{Fomichev_initial,berry2025rapid} and the implementation of real-time evolution under the Hamiltonian~\cite{roggero2020,waltersson2013}. While fault-tolerant quantum computers remain under development, it is interesting to identify models well suited to quantum computation and to estimate the resources needed to solve them. In this paper, we address how to construct scalable nuclear Hamiltonians and prepare initial states using the adaptive
derivative-assembled pseudo-Trotter variational quantum eigensolver (ADAPT-VQE)~\cite{peruzzo2014,grimsley2019}.

We want to pursue scalable quantum computations of nuclear systems. To understand the potential and limitations, let us consider the structure of the Hilbert spaces and the Hamiltonians acting on them. A system of $n_q$ qubits spans a Hilbert space of dimension $2^{n_q}$, enabling the representation of exponentially many quantum states. Similarly, the Hilbert space dimension of an $A$-body nuclear system also grows exponentially with the mass number $A$. This implies that quantum computers can encode such wave functions using a number of qubits that only grows linearly with $A$, providing an exponential advantage in terms of memory storage.

However, in a Gray-code encoding (where a matrix representation of the $A$-body Hamiltonian is used)~\cite{dimatteo2021,Siwach_nuclear_gray_encoding,singh2025advancing,singh2025,Kiss25_moment,weiss2025,chinnarasu2025} the number of Hamiltonian matrix elements increases faster than the Hilbert space dimension~\cite{maris2010b}. Thus, the translation of Hamiltonians into Gray-code operators scales exponentially with $A$. We see that in this framework quantum computing offers an exponential advantage for state representation while the construction and processing of the Hamiltonian operators remains exponentially costly. This is seen, e.g., in the quantum computing of $^6$Li of  Ref.~\cite{singh2025advancing}. There, the Hilbert space with angular momentum projection $J_z=0$ required for the computation of $J^\pi=0^+$ states had dimension $D=10$ (see Table XIII of that work for a complete list of $J^\pi=0^+$ states) and required four qubits, but the number of Gray-code operators was 134 (see Table XX of that work for a complete list of Gray-code operators), which exceeded the number $D(D+1)/2=55$ of independent matrix elements of a $D\times D$ real-symmetric Hamiltonian matrix.

Instead, we work in the framework of second quantization, as is also used in, e.g., Refs.~\cite{stetcu2022, perezobiol2023, Sarma_23_oxygen, Bhoy:2024uuh}. Then, $n_q$ qubits can represent the same number of single-particle states. Nuclear Hamiltonians consist of the kinetic energy (a one-body operator) and two- and three-body interactions. They are of the form
\begin{align}
    H &= \sum_{pq} \varepsilon_q^p \hat{a}_p^\dagger \hat{a}_q + {1\over 4}\sum_{pqrs} V^{pq}_{rs} \hat{a}_p^\dagger \hat{a}_q^\dagger \hat{a}_s\hat{a}_r \nonumber \\
    &+ {1\over 36}\sum_{pqrsuv} W^{pqr}_{suv} \hat{a}_p^\dagger \hat{a}_q^\dagger \hat{a}_r^\dagger \hat{a}_v\hat{a}_u\hat{a}_s \ .
\end{align}
We see that the Hamiltonian consists of ${O}(n_q^6)$ terms. Working in an axially symmetric framework (``$m$-scheme'') of the nuclear shell model~\cite{caurier2005} or in a momentum-space basis -- where the total momentum is conserved~\cite{hagen2013b} -- reduces this number to about ${O}(n_q^5)$ and ${O}(n_q^3)$, respectively. While this is manageable in classical computing, such a scaling poses a significant effort on present-day quantum devices and might preclude quantum advantages~\cite{hoefler2023}. As an example we again consider the quantum computation of $^6$Li, this time as performed in Ref.~\cite{kiss2022}. That approach used $n_q=12$ single-particle states (and qubits) but the Hamiltonian consisted of about 1000 Pauli terms. Hamiltonians in larger valence spaces that would be intractable on classical computers, such as the $pfsdg$~shell, easily have several hundred thousand Pauli terms. Clearly, one has to exploit the short range of the nuclear interaction. Using a lattice in position space~\cite{lee2009,lahde2019,watson2023,lee2025}, the number of Hamiltonian terms is of order ${O}(n_q)$; it only grows linearly with the number of lattice sites because the nuclear interaction is short ranged. 

In this work, we use such a single-particle basis and build on the recent works~\cite{roggero2018,roggero2020,baroni2022,watson2023} regarding quantum computing of nuclei on lattices. The works~\cite{roggero2018,roggero2020,baroni2022}  used two-dimensional lattices for computations of two-nucleon systems (and three nucleon systems where one nucleon is static), while Ref.~\cite{watson2023} studied resource requirements for lattice Hamiltonians. Here, we present quantum computations of $A=2,3$ dynamical nucleon systems on three-dimensional lattices.

This paper is organized as follows: Section~\ref{sec:Model} introduces the lattice model space and Hamiltonian. In Sec.~\ref{sec:QuantumComputing} we give details on the quantum computing algorithms used in Sec.~\ref{sec:Results} to compute the structure of the deuteron and $^3$He. We conclude with a summary and perspectives in Sec.~\ref{sec:Outro}. Some technical details are presented in the appendixes.

\section{Lattice Hamiltonian}
\label{sec:Model}

We have a three-dimensional lattice of 
\begin{equation}
    n=L^3
\end{equation}
sites and use cubic boundary conditions. 
We label lattice sites using the indices $0,1,2,\ldots,n-1$. In general we need to employ $n_q=4n$ single-particle states because of two isospin and two spin projections. For the quantum computations we use $n_q$ qubits. 
We use a Hamiltonian from pionless effective field theory (EFT)~\cite{bedaque2002,Hammer:2019poc}. At leading order, the interaction consists of a two-body contact and a three-body contact~\cite{bedaque1999}. 
The Hamiltonian is
\begin{align}
\label{ham}
    \hat{H} &= \sum_{\langle\mathbf{l},\mathbf{l}'\rangle}\sum_{\tau s} T_{\mathbf{l}'}^{\mathbf{l}} \hat{a}_{\mathbf{l}\tau s}^\dagger \hat{a}_{\mathbf{l}'\tau s} \nonumber\\
    &+ {V\over 2}\sum_{\mathbf{l}}\sum_{ss'\tau\tau'} \hat{a}_{\mathbf{l}\tau s}^\dagger \hat{a}_{\mathbf{l}\tau' s'}^\dagger \hat{a}_{\mathbf{l}\tau' s'}\hat{a}_{\mathbf{l}\tau s} \\
    &+ W\sum_{\mathbf{l}}\sum_{\tau s} \hat{a}_{\mathbf{l}\tau\uparrow}^\dagger \hat{a}_{\mathbf{l}\tau\downarrow}^\dagger \hat{a}_{\mathbf{l}-\tau s}^\dagger \hat{a}_{\mathbf{l}-\tau s}\hat{a}_{\mathbf{l}\tau \downarrow}\hat{a}_{\mathbf{l}\tau \uparrow} \nonumber \ .
\end{align}
Here, the operator $\hat{a}_{\mathbf{l}\tau s'}$ creates a nucleon with isospin projection $\tau$ and spin projection $s$ on the lattice site denoted by the integer lattice vector $\mathbf{l}=(l_x, l_y, l_z)$. The notation $\langle\mathbf{l},\mathbf{l}'\rangle$ indicates that $\mathbf{l}$ and $\mathbf{l'}$ are either the same site or nearest neighbors. The matrix elements of the kinetic energy are denoted as 
$T_{\mathbf{l}'}^{\mathbf{l}}$. We use
\begin{align}
\label{Tkinmatele}
T_{\mathbf{l}'}^{\mathbf{l}} = -\frac{\hbar^2}{2ma^2}\sum_{i = x,y,z}\left(\delta_{\mathbf{l}'}^{\mathbf{l}-\mathbf{e}_i}    -2\delta_{\mathbf{l}'}^{\mathbf{l}} + \delta_{\mathbf{l}'}^{\mathbf{l}+\mathbf{e}_i} \right) , 
\end{align}
where $m$ is the nucleon mass, $a$ is the lattice spacing, and $\mathbf{e}_i$ is a unit vector in the direction $i=x,y,z$. This is the leading order approximation of the Laplacian on the lattice and corrections are of order ${O}(a^2)$ relative to the leading approximation~(\ref{Tkinmatele}).
We use
\begin{equation}
    V = \frac{\hbar^2v}{2ma^2}
\end{equation}
and
\begin{equation}
    W = \frac{\hbar^2w}{2ma^2} \ , 
\end{equation}
where $v$ and $w$ are dimensionless couplings.
Inspection shows that the two-body interaction of the Hamiltonian~(\ref{ham}) is spin-isospin invariant and thereby displays Wigner's SU(4) symmetry.

The Hamiltonian~(\ref{ham}) implements pionless EFT at leading order. That theory is a low-energy effective field theory of quantum chromodynamics~\cite{bedaque2002} and has been used to describe few-nucleon systems~\cite{bedaque1999,platter2005,kirscher2010,lensky2016,barnea2015}. It reflects that nuclear interactions -- in their crudest approximation -- are short ranged and nearly spin-isospin symmetric.

Let us also account for the resource requirements. For nuclei with mass number $A\ge 4$, a lattice with $L^3$ sites and four spin-isospin states per site requires $n_q=4L^3$ qubits. For the Hamiltonian~(\ref{ham}), the kinetic energy consists of seven Pauli terms per lattice site and spin-isospin state (six to hop away and one to stay on the site), making a total of $7 n_q$ Pauli terms. The two-body contact has six (four choose two) Pauli terms  per lattice site, and the three-body contact has four (four choose three) Pauli terms per lattice site. Thus, there are about $10n_q$ Pauli terms in the Hamiltonian. The proportionality between the number of Pauli terms and the number of qubits is a highlight of short-ranged Hamiltonians. For the $A=2,3$ nuclei these numbers are further reduced. The deuteron only requires $n_q=2L^3$ qubits, and the $A=3$ system only $n_q=3L^3$ qubits for our on-site SU(4) symmetric interaction. 

The Hamiltonian~(\ref{ham}) may be systematically improved for more precise computations of nuclei using techniques from effective field theory~\cite{Hammer:2019poc}. More accurate lattice Hamiltonians can be found in Refs.~\cite{lee2009,elhatisari2016,elhatisari2024,lee2025} and we briefly mention some of their features.  Improved approximations of the Laplacian couple lattice sites up to next-to-next-to-nearest neighbors~\cite{epelbaum2010}. 
A more accurate description of nuclei requires higher-order contributions, nonlocal interactions~\cite{elhatisari2016} and interactions from pion-exchange~\cite{epelbaum2009}, which both introduce spin-orbit and tensor interactions.
Higher-order contact interactions have a short, but finite range. Pion-exchange has the range of the inverse pion mass $m_\pi^{-1}\approx 1.5$~fm and  will predominantly couple nearest and next-to-nearest neighbors for typically used lattice spacings $a\sim1.5$--$2$~fm.

Crucially, any such refinement involving finite-range interactions does not change the fact that the number of matrix elements scales linearly in the basis size $n_q$.
This also applies to finite-range approximations of the Laplacian.\footnote{
The exact kinetic energy is local in Fourier space and thus completely nonlocal in position space, which would give a scaling of $O(n_q^2)$.
In the case that finite-range approximations to the Laplacian are insufficient for the desired precision, this scaling would still be favorable to the $O(n_q^3)$--$O(n_q^5)$ scaling in other bases.}
This can be easily understood. The finite range allows particles to interact in a small volume, but this volume is independent of the total volume of the simulation. The result is that the scaling is still linear in the basis size, simply with a different prefactor.
This prefactor may be large, and this will accordingly increase the computational cost of the simulation, but for computations in large bases it is still very favorable compared with $O(n_q^3)$ and $O(n_q^5)$ scaling in other bases.
(We recall that state of the art ab initio calculations of nuclei on classical computers employ basis sizes using 2000--5000 single-particle states~\cite{hu2022,elhatisari2024}.) It is clear that no other single-particle basis offers a more favorable scaling regarding the number of Hamiltonian terms.

The Coulomb interaction is long range and delivers small, but important contributions for nuclear structure. In pionless EFT, it contributes at next-to-leading order for ground-state energies~\cite{konig2016,Konig:2016utl,Kirscher:2015zoa} and may be treated in perturbation theory~\cite{epelbaum2010} following a nonperturbative ground-state solution.
Such a perturbative treatment would also be an option in quantum computing contexts,
with the obvious challenge that the number of nonzero matrix elements scales like $O(n_q^2)$ (due to being long range and local).
Efficiently evaluating this contribution either via classical postprocessing or via appropriate quantum algorithms will be an important problem for future work.

We adjust the coupling constants such that exact diagonalizations on sufficiently large lattices semiquantitatively reproduce the binding energies of light nuclei (see Appendix~\ref{app:ham} for details). For the quantum computations in this work, we use $L=2$, $a = 2.0\:\text{fm}$, $v = -9.0$, and $w=10.8$. On such a small lattice, finite-size corrections are substantial and the two- and three-body binding energies are about 12.9 and 29.5~MeV, respectively.  

The Hamiltonian~(\ref{ham}) conserves isospin, total spin, and their projections, as well as the parity and is invariant under discrete lattice translations.
The discrete translation symmetry implies that all eigenstates of the Hamiltonian are products of intrinsic states and center-of-mass states. The intrinsic eigenstates have vanishing kinetic energy in the three degrees of freedom of the center of mass. ``Spurious'' eigenstates have a finite energy in the center of mass.

Let us briefly comment on the special structure of the Hamiltonian~(\ref{ham}) where the potential is diagonal in the chosen lattice basis and the kinetic energy is off-diagonal. The recent study~\cite{Rothman:2025uza} revealed that Hartree Fock yields about 81\% of the ground-state energy for $^3$He computed with a lattice spacing $a=2$~fm (as used in this work). Furthermore, the coupled-cluster method with single and double excitations~\cite{bartlett2007,hagen2014} or the in-medium similarity renormalization group~\cite{tsukiyama2011,hergert2016} yield little improvement over the Hartree-Fock energy. Our results below meet benchmarks from exact diagonalization, and this shows that three-particle--three-hole excitations are fully captured (for the $^3$He nucleus) in our quantum computations.  
In summary, Hartree Fock captures a large portion of the ground-state energy because the kinetic energy is nonlocal while two- and three-body interactions are on site. However, obtaining the exact energy requires exact or high-order many-body expansion methods.

\section{Quantum computation}
\label{sec:QuantumComputing}

The variational quantum eigensolver (VQE) has emerged as a promising approach for near-term quantum simulations, particularly for ground state problems in quantum many-body physics~\cite{peruzzo2014,tilly2022}. The VQE approach combines a parametrized quantum circuit, known as an ansatz, with a classical optimizer to minimize the expectation value of the Hamiltonian. While conceptually appealing and implementable on current noisy hardware, standard VQE approaches face severe practical limitations. These include a high number of measurements required to estimate expectation values with sufficient precision and the possibility of encountering barren plateaus, i.e., regions of vanishing gradients, in the optimization landscape~\cite{ragone2024, Larocca2025BarrenPlateaus}.

To mitigate the trainability issue, we employ the adaptive derivative-assembled pseudo-Trotter VQE (ADAPT-VQE) algorithm~\cite{grimsley2019,tang2021}, an iterative variant of VQE that dynamically constructs the variational ansatz during the optimization process. ADAPT-VQE adaptively grows the circuit by selecting the most impactful operators from a predefined operator pool. At each iteration, the operator that gives the steepest descent in energy is appended to the ansatz, ensuring that only the most relevant excitations are included. The advantage of ADAPT-VQE is that it is expected to deliver the most shallow circuit that can be expressed from the pool and at the same time accelerates the convergence of the variational optimization. ADAPT-VQE has been applied in nuclear physics, albeit mainly within the shell-model framework~\cite{kiss2022,perezobiol2023}. In this work, we focus instead on a scalable lattice formulation, which provides a complementary setting for exploring ADAPT-VQE.

A crucial component of the ADAPT-VQE framework is the design of the operator pool, which must balance physical expressiveness and circuit efficiency. For our purposes, we construct the operator pool using the kinetic terms of the Hamiltonian~(\ref{ham}) along with correlated two-body hopping operators. These operators are selected to preserve spin, isospin, and particle number. This physically informed construction helps reduce the search space and accelerates convergence while ensuring that the resulting variational ansatz remains compact and interpretable.
One can imagine further tailoring the operator pool the problem of interest to efficiently capture the expected dominant states.
Such tailoring, if done correctly, can lead to shallower circuits in VQE computations, as was shown in Ref.~\cite{Bhoy:2024uuh} for shell-model quantum computations of $^{58}$Ni.
Developing an expressive, but general operator pool is clearly important for generally parametrizing the wave function while minimizing the circuit depth obtained using variational algorithms.

\subsection{System details}

In this work, we perform quantum computations of the deuteron and $^3$He.
The deuteron is the spin $S=1$ bound state of a proton and a neutron. This simplifies the calculation as follows: We choose an initial state 
\begin{equation}
\label{2H-initial}
    |\phi_0\rangle = \hat{a}_{\mathbf{l}{-\frac{1}{2}} \downarrow}^\dagger \hat{a}_{\mathbf{l}{+\frac{1}{2}} \downarrow}^\dagger |\slashed{0}\rangle\,,
\end{equation}
where the proton and neutron both occupy the same lattice site $\mathbf{l}$ and have identical spin projections; to be specific, we choose each nucleon as spin down. The vacuum state (i.e., the empty lattice) is $|\slashed{0}\rangle$. The initial state clearly has total spin $S=1$ (and projection $S_z=-1$). As the Hamiltonian~(\ref{ham}) preserves the spin projection, one only needs $n_q=2L^3$ qubits, i.e., both the proton and the neutron will remain in states with spins down. Thus, the simplest computation on a lattice with $L=2$ requires 16 qubits.

The $^3$He nucleus is the spin $S=1/2$ bound state of two protons and a neutron. This simplifies the calculation as follows. We choose an initial state 
\begin{equation}
\label{3He-initial}
    |\phi_0\rangle = \hat{a}_{\mathbf{l}{-\frac{1}{2}} \downarrow}^\dagger
    \hat{a}_{\mathbf{l}{-\frac{1}{2}} \uparrow}^\dagger
    \hat{a}_{\mathbf{l}{+\frac{1}{2}} \downarrow}^\dagger |\slashed{0}\rangle \,,
\end{equation}
where all nucleons occupy the same lattice $\mathbf{l}$. We choose the total spin projection $S_z=-1/2$. This again simplifies the computation because the neutron will stay in a spin down state because of our SU(4) symmetric interaction. Thus, the quantum computation of the $^3$He nucleus with the Hamiltonian~(\ref{ham}) requires $n_q=3L^3$ qubits; for $L=2$ this is $n_q=24$. We see that the $A=2,3$-body systems require smaller numbers of qubits.  The $\alpha$ particle and heavier nuclei will require $n_q=4L^3$ qubits.

\subsection{ADAPT-VQE calculations}

Our ADAPT-VQE computations are based on a unitary coupled cluster ansatz~\cite{kutzelnigg1983,taube2006,evangelista2019,anand2022} for the variational state
\begin{align}
    |\vec{\theta}\rangle = \prod_{\alpha=1}^{N_{\rm e}} \exp{\left(\sum_{\beta=1}^{N_{\rm p}} \theta_{\alpha\beta}\hat{A}_{\alpha\beta}\right)}|\phi_0\rangle \,.
    \label{eq:ADAPTAnsatz}
\end{align}
Here $\vec{\theta}=\{\theta_{\alpha\beta}\}$ is a set of real variational parameters, and $\hat{A}_{\alpha\beta}$ are anti-Hermitian operators. The number of operators $N_{\rm p}=10$ is taken from a larger pool of operators which is discussed below, and $N_{\rm e}$ is the number of exponentials.

A central ingredient of the ADAPT-VQE approach is the construction of the operator pool $\hat{A}_{\alpha\beta}$, which defines the variational directions available to the ansatz. For a general Hamiltonian expressed as
\begin{equation}
\label{Hsum}
    \hat{H} = \sum_\sigma v_\sigma \hat{h}_\sigma\,,
\end{equation}
with $\hat{h}_\sigma$ denoting products of Pauli operators, a typical choice is to include anti-Hermitian generators of the form $\hat{A}_{\alpha\beta} \in \{ i \hat{h}_\sigma \}$ taken directly from the Hamiltonian~\cite{stetcu2022,kiss2022,perezobiol2023}.

In our case, however, this standard strategy proves inadequate. The two-body terms in our Hamiltonian~(\ref{ham}) act locally on single lattice sites and do not generate hopping between neighboring sites. As a result, they fail to significantly reduce the energy when applied to the initial reference state $|\phi_0\rangle$. Moreover, the one-body terms when exponentiated merely induce a transformation of the single-particle basis and are therefore incapable of generating the necessary many-body correlations. This motivates the inclusion of additional nonlocal two-body operators—specifically correlated two-body hopping terms—that respect spin, isospin, and particle number conservation.

A second and important consideration concerns the algebraic structure of the operators used to generate the variational ansatz. While conventional approaches rely on unitary transformations generated by exponentiating purely imaginary anti-Hermitian operators (such as $i\hat{h}_\sigma$), our Hamiltonian is real and symmetric rather than complex Hermitian. In such a case, an orthogonal transformation—sufficient to preserve the structure of a real-valued wave function—is adequate to reach the ground state from $|\phi_0\rangle$. Since the orthogonal group is a real subgroup of the unitary group, such a restriction to orthogonal transformations can simplify the optimization while still accessing the relevant part of Hilbert space.

This distinction happens to be essential in our setting. For example, moving nucleons between sites using purely imaginary generators does not lead to energy reduction due to the local nature of the Hamiltonian. In contrast, orthogonal transformations based on the exponentiation of real-antisymmetric generators enable effective exploration of the relevant variational manifold. 
To this end, we construct our operator pool from real antisymmetric matrices, which generate orthogonal transformations after exponentiation. 
A detailed discussion of our operator construction is provided in Appendix~\ref{app:operators}.

The intrinsic ground state must be invariant under discrete lattice translations. While it is in principle possible to generate initial states with this property (see Appendix~\ref{app:trans} for details), we refrain from doing so for the following reasons. First, the unitary operator that yields a translationally invariant state when acting on $|\phi_0\rangle$ will be approximated, e.g., via Trotterization. Thus, it cannot  be implemented exactly. Second, the operators in the pool used by ADAPT-VQE break translational invariance. While one could contemplate using only translationally invariant combinations of operators, this would complicate the variation of the state. Third, the energy carried by spurious states decreases like $A^{-1}$ with increasing mass number and thus becomes less and less important in heavier nuclei. Finally we mention that one could also consider intrinsic Hamiltonians where the kinetic energy of the center of mass $T_{\rm CoM}$ is subtracted from Eq.~(\ref{ham}). Then the number of Pauli terms in the Hamiltonian would increase from ${O}(n_q)$ to ${O}(n_q^2)$, because the $T_{\rm CoM}$ is a two-body operator in single-particle momenta.

In our calculations, we need to identify the most efficient operators $\hat{A}_{\alpha\beta}$, optimize the variational parameters $\theta_{\alpha\beta}$, and truncate our ansatz at some fixed number of exponentials $N_\mathrm{e}$. We approach this problem sequentially, where in each optimization epoch we expand our ansatz and perform a subsequent optimization of the added variational parameters. We start from our initial state $\ket{\phi_0}$. For $\alpha = 1$, we select the $\beta = 1, \dots,N_\mathrm{p}$ operators $\hat{A}_{1\beta}$ that have the largest energy gradients around $\theta_{1\beta} = 0$. The selected variational parameters $\theta_{1\beta}$ are then optimized via gradient descent, with a maximum of 100 optimization iterations. The resulting optimized state serves as the new reference for the subsequent iteration, in which the procedure is repeated for $\alpha = 2$, and so forth. This iterative process continues, with each step involving the identification and optimization of additional operators.

As the variational state progressively approaches the ground state, the magnitudes of the energy gradients diminish, and the decrease in the Hamiltonian expectation value resulting from the inclusion of further exponential terms becomes increasingly marginal. In the limit of vanishing gradients, no further energy reduction can be achieved through the addition of operators in the ansatz of Eq.~\eqref{eq:ADAPTAnsatz}.

\subsection{Sample complexity}
\label{sec:SampleComplexity}

The dominant cost in variational quantum algorithms typically arises from the number of measurements required in the quantum-classical feedback loop \cite{Mazzola_shot_noise}. While the number of optimization iterations needed for convergence is difficult to predict, we focus here on the sample complexity required to estimate the expectation value of the energy to within a fixed standard deviation \( \varepsilon \).

Consider a Hamiltonian of the form \eqref{Hsum}. The goal is to estimate the expectation value \( \langle \hat{H} \rangle \) such that the standard deviation is at most \( \varepsilon \).

If \( s_\sigma \) denotes the number of measurements (shots) allocated to \( \hat{h}_\sigma \) and we assume the variance \( \text{Var}(\hat{h}_\sigma) \leq 1 \), then the total variance in the energy estimate satisfies the bound
\begin{equation}
\begin{split}
    \text{Var}[\langle \hat{H} \rangle] &= \langle \psi|\hat{H}^2|\psi\rangle - \langle \psi|\hat{H}|\psi\rangle^2 \\ 
    &\leq \sum_\sigma \text{Var}[\langle v_\sigma h_\sigma \rangle] \leq \sum_\sigma \frac{v_\sigma^2}{s_\sigma}\,.
    \end{split}
\end{equation}
To understand the challenges faced by the VQE, it is important to note that the variance of the energy estimator remains strictly positive—even when evaluated on the exact ground state. This arises because the expectation value of the Hamiltonian cannot be computed directly---at least without using ancilla qubits---but must instead be decomposed into a sum of measurable terms. This behavior stands in contrast to traditional Monte Carlo methods, where the variance can in principle vanish when sampling from the exact ground state. 

The optimal allocation of measurements that minimizes the total number of samples \( \sum_\sigma s_\sigma \) subject to the constraint \( \text{Var}[\langle \hat{H} \rangle] \leq \varepsilon^2 \) can be derived using Lagrange multipliers.

The nuclear lattice Hamiltonian can be rewritten as
\begin{equation}
    \hat{H} - a \mathbb{1} = \hat{T} + \hat{V}\,,
\end{equation}
where \( a \in \mathbb{R} \) is a constant, \( \hat{T} \) contains noncommuting terms, and \( \hat{V} \) consists of mutually commuting terms that are simultaneously diagonalizable.

Let \( \mathcal{T} \) and \( \mathcal{V} \) denote the index sets corresponding to the noncommuting and commuting terms in \( \hat{T} \) and \( \hat{V} \), respectively. In the following, we make the reasonable assumption that individual estimators from commuting observables remain uncorrelated when measured together. This allows us to group all the terms in \( \mathcal{V} \) since they can measured simultaneously, resulting in a drastic shot reduction. The variance of the total-energy estimate can be expressed as
\begin{equation}
    \text{Var}[\langle \hat{H} \rangle] \approx \frac{1}{s_V} \sum_{\sigma \in \mathcal{V}} v_\sigma^2 + \sum_{\sigma \in \mathcal{T}} \frac{v_\sigma^2}{s_\sigma}\,,
\end{equation}
where \( s_V \) is the shared number of measurements used to estimate all commuting terms, and \( s_\sigma \) is the number of measurements allocated individually to each noncommuting term.

The optimal allocation minimizing the total number of measurements under this constraint is given by
\begin{align}
    s_V &= \frac{1}{\varepsilon^2} \frac{\sum_{\sigma \in \mathcal{V}} v_\sigma^2}{\| \hat{H} \|_2^2}\,, \\
    s_\sigma &= \frac{1}{\varepsilon^2} \frac{v_\sigma^2}{\| \hat{H} \|_2^2} \quad \text{for } \sigma \in \mathcal{T}\,,
\end{align}
where $\| \hat{H} \|_2^2$ denotes the Frobenius norm.

This strategy ensures the most efficient distribution of samples for estimating \( \langle \hat{H} \rangle \) up to error \( \varepsilon \) by allocating a shared shot count to the commuting group \( \mathcal{V} \) and individual shot counts to the noncommuting terms in \( \mathcal{T} \).

\begin{figure}[t!]
    \centering
    \includegraphics{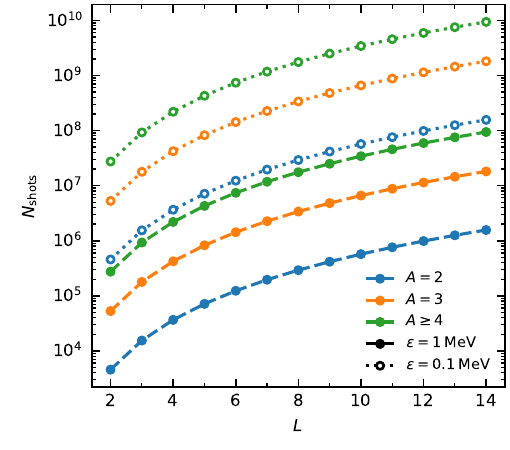}
    \caption{Scaling of the number of samples \Nshots{} required to reach a given precision in the lattice extent $L$ and the system size $A$. Going from $\varepsilon = 1~\mathrm{MeV}$ (filled circles) to 0.1~MeV (open circles) requires two orders of magnitude more samples. For $A=2,3$ we are able to make simplifications such that ADAPT-VQE computations only require $n_q = 2 L^3, 3L^3$ qubits, respectively, rather than the general $n_q = 4L^3$. }
    \label{fig:shot_scaling}
\end{figure}

Based on our selected dimensionless couplings $v= -9.0$, $w=10.8$ and our lattice spacing $a = 2.0\:\text{fm}$, we compute the shot budget required to estimate the energy within 1 and $0.1$~MeV as a function of the lattice extent $L$ and system size $A$. 
The results, shown in Fig.~\ref{fig:shot_scaling}, show that the number of shots required strongly depends on the desired precision.
Going from $\varepsilon = 1$ to $0.1\:\mathrm{MeV}$ requires approximately 100 times as many samples.
For increasing lattice sizes the required number of samples scales like the volume $L^3$.
For lattice sizes $L=2$--4, the number of shots required to reach $\varepsilon = 1\:\mathrm{MeV}$ is on the order of $10^4$--$10^6$ depending on the system, while increasing to $L=10$ increases this by another order of magnitude.
Quantum computations with $\Nshots \sim 10^6$ are expensive, but possible on current hardware,
and this is lower than the number of shots required for quantum chemistry applications, where as many as $10^{10}$ shots may be required to reach chemical accuracy in the case of, e.g., ethanol~\cite{Measurement-complexity-VQE}.
We note that those resource estimates are only for the resources required to estimate the energy of a given state and thus do not entail the optimization cost.

\section{Results from quantum simulations}
\label{sec:Results}

\begin{figure*}
    \centering
    \includegraphics{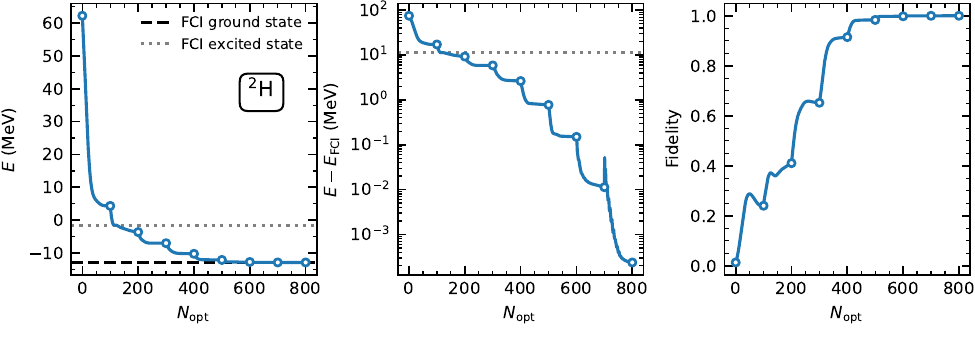}
    \caption{
        Ground state energies and state fidelities obtained using ADAPT-VQE to solve the deuteron. Results are shown as a function of the number of optimization steps \nopt{}. We optimize the parameters $\theta_{\alpha\beta}$ for a selection of 10 operators $\hat{A}_{\alpha\beta}$ in each epoch $\alpha$, where the start of an epoch and the selection of a new set of operators is indicated by the open points. In each epoch, we use at most 100 optimization iterations to identify an optimal set of parameters. The exact ground state energy and the threefold degenerate first excited-state energy for our Hamiltonian, computed via exact diagonalization, are indicated as dashed black and dotted gray lines, respectively.
        \label{fig:h2_no_noise}
    }
\end{figure*}

We simulate quantum computations of the deuteron and $^3$He on classical computers with the \textsc{PennyLane} library~\cite{Bergholm2018pennylane}, using \textsc{jax}~\cite{jax2018github} to evaluate energy gradients for the operators $\hat{A}_{\alpha\beta}$ and \textsc{Optax}~\cite{deepmind2020jax} to optimize the parameters $\theta_{\alpha\beta}$ in our ADAPT-VQE ansatz.
We perform both exact simulations, where energy gradients and Hamiltonian expectation values computed from our simulated circuits are evaluated exactly, and noisy simulations using a finite number of measurements \Nshots{} of the simulated circuits.

\subsection{Simulations without noise}

Our exact simulation of the deuteron is summarized in Fig.~\ref{fig:h2_no_noise}. We show the ground-state energy, the energy difference with the exact ground state energy $E_\mathrm{FCI}$ as computed from an exact diagonalization [or full configuration interaction (FCI)], and the fidelity of our state $|\vec{\theta}\rangle$ as a function of the number of optimization steps.

Our initial state has an energy of 62.206~MeV, far from the exact ground state energy of $-12.874$~MeV. The corresponding fidelity with the exact ground state $\ket{\psi}$, computed as $|\braket{\psi}{\phi_0}|^2$, is very low, specifically 0.013. In this case, our initial state is clearly too compact to quantitatively describe the weakly bound deuteron, leading to the large energy expectation value and low fidelity. Additionally, our initial state is not translationally invariant; the construction of a symmetry preserving state, as discussed in Appendix~\ref{app:trans}, would require the preparation of a complicated linear combination of our initial state on all sites of the lattice. The exact ground state is of course translationally invariant, which is another contributing factor to the low fidelity of $\ket{\phi_0}$ with the exact ground state.

In our first two epochs, we see a substantial decrease in the ground-state energy, decreasing below the threefold degenerate (and spurious) first excited state by the end of the second epoch. During this phase of our calculation, the fidelity increases substantially, but not quite monotonically because the optimizer spends some time exploring the local saddle points in parameter space associated with the excited states. In later epochs, as our ADAPT-VQE ansatz grows in terms of the number of exponentials \Nexp{}, the energy continues to decrease, converging to within roughly 1~MeV of the exact ground state energy in 500 optimization steps with $\Nexp = 5$. At this point, the fidelity is also very close to 1. Further expanding and optimizing our ansatz continues to improve our state to the point where reproduction of the exact ground state and ground state energy could basically be considered exact. 

\begin{figure*}
    \centering
    \includegraphics{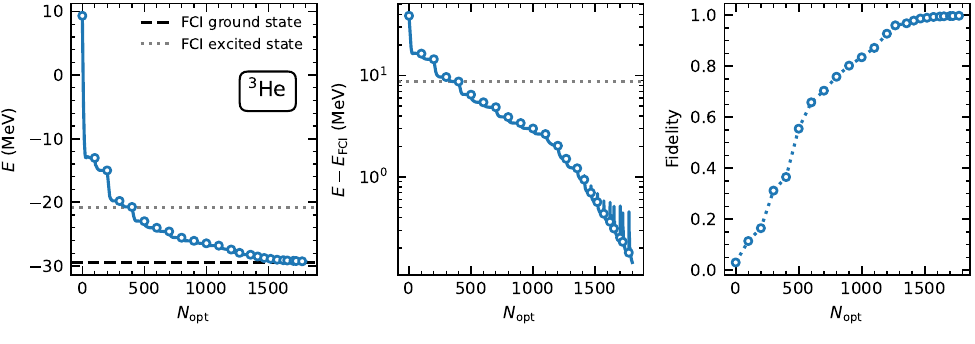}
    \caption{
        Same as Fig.~\ref{fig:h2_no_noise}, but for $^{3}\mathrm{He}$. State fidelities were only computed at the end of each epoch, so the dashed lines in the right panel are only intended to guide the eye.
        \label{fig:he3_no_noise}
    }
\end{figure*}

Figure~\ref{fig:he3_no_noise} shows similar results for our exact simulation of $^3$He. Our initial state has an energy of 9.331~MeV, considerable closer to the exact ground state energy of $-29.468$~MeV than in the case of the deuteron. Still the initial state lies far above the first excited state and has a very small fidelity of 0.030, which we attribute to missing translational invariance and the too-compact structure of the initial state. However, the fidelity of the initial state for $^3$He is larger than for the deuteron, presumably because the former actually is more compact than the latter.  

We find that our calculation systematically converges to the exact ground state energy and achieves essentially perfect fidelity with the exact ground state. The number of optimization steps required is larger than for the deuteron, but even after eight epochs, we already predict an energy closer to the ground state energy than the first excited state energy. We partially attribute the longer optimization time to the larger pool of potential operators for our ADAPT-VQE ansatz due to the larger basis size ($n_q = 3L^3$ rather than $2L^3$). However, it is important to note that we do not include any additional three-body operators in our ansatz for our $^3$He calculations. This is crucial, because we are able to efficiently capture the essential correlations in the three-body system using an ansatz consisting of only one- and two-body operators, suggesting that such an approach may also scale well to heavier systems. This is also intuitive, as in classical many-body methods the truncation at the normal-ordered two-body level has been demonstrated to be a very effective, controlled approximation~\cite{hagen2007, soma2014, hergert2020, heinz2025}.

We can contrast this to the work of \textcite{perezobiol2023}, where they performed quantum shell-model computations of $^{20,22,24}$Ne of four, six, and eight interacting nucleons, respectively, using 24 qubits.
Their ADAPT-VQE simulations reached precisions of 2\% after including 167--345 layers in the ADAPT-VQE ansatz.
Our calculation of $^{3}$He reaches the same precision at 30 layers, clearly demonstrating that a simpler ansatz with a significantly smaller circuit depth is sufficient in our case.
This is due to the simplicity of the Hamiltonian and the natural expression of physically relevant operators (e.g., nearest-neighbor hopping operators) on the position-space lattice.
A more expressive operator pool could potentially even further simplify the required ansatz and reduce the circuit depth, as found for shell-model calculations in Ref.~\cite{Bhoy:2024uuh}.

We note that in Figs.~\ref{fig:h2_no_noise} and~\ref{fig:he3_no_noise} the difference to the exact ground-state energy decreases approximately exponentially in the number of epochs (or equivalently the number of exponentials in our ansatz). This could potentially be used to estimate the complexity of the ansatz required to reach a specific precision based on computations using only a small number of exponentials. There is also a nontrivial role played in the optimization by the energy of the first excited state. We see that the exponential decrease in the difference really kicks in after our state has been optimized below the first excited state. Additionally, in calculations of $^3$He without three-body interactions, where the ground and excited states are only separated by 1.954~MeV, we found that the optimization stalls for several epochs at the first excited state due to vanishing energy gradients. This suggests that a high fidelity with the exact ground state and a sizable energy difference between ground and excited states are beneficial to our simulations. In this case, our first excited states are center-of-mass excitations with the same intrinsic energy but nonzero total momentum, which are uninteresting for predictions of nuclear ground state properties. It remains an interesting question for future work how to handle these excitations in our simulations such that state optimization is as efficient as possible.

\subsection{Simulations with measurement noise}

\begin{figure}
    \centering
    \includegraphics{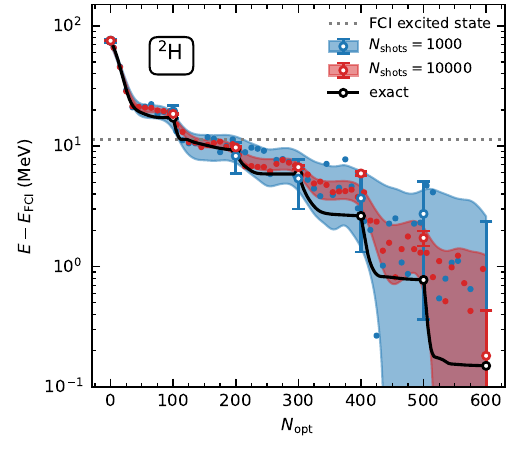}
    \caption{
        \label{fig:deuteron_noisy}
        Differences of ground state energies of the deuteron with the exact ground state energy $E_\mathrm{FCI}$ of the deuteron as computed in ADAPT-VQE simulations including measurement noise for $\Nshots = 1000$ (blue), 10000 (red) and without measurement noise (black). Results are shown as a function of the number of optimization steps \nopt{}.
        The open points with error bars are results at the end of each optimization epoch, with uncertainties estimated as $E_\mathrm{corr} / \sqrt{\Nshots}$ based on the correlation energy and the number of shots.
        For clarity, we only show 10\% of the evaluated energies obtained during the optimization as small, filled-in points. The estimated uncertainty on the energy during the optimization is indicated by the band.
    }
\end{figure}

In the simulations so far, our quantum circuits are evaluated exactly, giving exact energy expectation values and gradients. In Fig.~\ref{fig:deuteron_noisy}, we explore the impact of measurement uncertainties by simulating the ADAPT-VQE calculation of the deuteron with a finite number of evaluations \Nshots{} per energy gradient and energy expectation value evaluation. We estimate the resulting uncertainty simply as $E_\mathrm{corr} / \sqrt{\Nshots}$ based on the correlation energy, the energy difference between the fully correlated state and our initial state $E_\mathrm{corr} = 75.078$~MeV, and the number of samples \Nshots{}. This uncertainty is about 2.5~MeV and 750~keV for $\Nshots{} = 1000$ and 10000, respectively, which sets the best precision we can expect to achieve in our calculations.

We compare $\Nshots{} = 1000$ (blue), 10000 (red) with the exact simulation discussed above in black. We see that in the first few epochs, the sampling uncertainty does not pose a significant challenge for the optimizer and the calculations with $\Nshots{} = 1000$, 10000 both closely follow the exact calculation. Once we have converged within about 5--10~MeV of the ground state energy, the sampling uncertainty slows the optimization down slightly, but in both cases the ground state energy continues to systematically decrease. By the completion of the sixth epoch, we see that the intrinsic uncertainty from the finite number of samples limits the further optimization of our state. At this point the energy difference to the exact state is of roughly the same size as the estimated sampling uncertainty on the energy, and so additional epochs would only be able to improve the state if a larger number of samples would be used. We do not explore such adaptive optimization strategies in this work. 

The results from Fig.~\ref{fig:deuteron_noisy} are consistent with the estimates presented in Fig.~\ref{fig:shot_scaling} because the $A=2$ system indeed requires on the order of $10^4$ shots to get a 1~MeV accuracy. 

\subsection{Practical considerations and perspectives}

\begin{figure}
    \centering
    \includegraphics{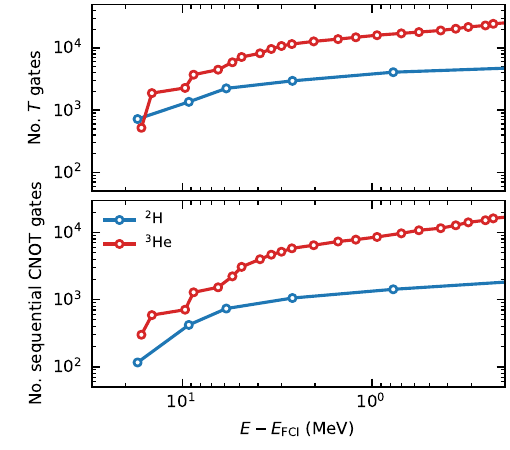}
    \caption{
        \label{fig:cnots}
        Number of \(T\) (top) and sequential CNOT (bottom) gates in the circuit for our ADAPT-VQE ansatz as the number of exponentials \Nexp{} is increased for the deuteron and $^3$He.
        As the number of exponentials is increased, the ansatz grows more complicated and the circuit grows due to the added operators.
        At the same time, the error to the exact ground state energy ($E-E_\mathrm{FCI}$ plotted on the $x$ axis) is systematically decreased.
        We only show points with $\Nexp \geq 1$, where the number of \(T\) and CNOT gates is larger than zero.
    }
\end{figure}

Quantum computations are limited by different resources depending on the hardware regime: noisy intermediate-scale quantum (NISQ) devices~\cite{preskill2018} and fault-tolerant quantum computers. 
On NISQ devices,  the primary constraint is the number of two-qubit gates, typically quantified by the number of CNOT operations. Due to the accumulation of noise, only a limited number of such gates can be reliably executed before measurement outcomes become dominated by error. This imposes an upper bound on circuit depth. Additionally, repeated circuit executions are required to estimate expectation values and gradients accurately, leading to a further constraint due to the total number of measurement shots.
In the fault-tolerant regime, where quantum error correction enables arbitrarily deep circuits, the dominant resource becomes the number of non-Clifford gates, especially the \(T\) gate. While Clifford operations can be implemented efficiently within error-corrected codes, \(T\) gates require costly procedures such as magic state distillation. Furthermore, amplitude amplification and other quantum techniques allow for a reduction in the number of required samples for estimating expectation values.

With these two practical concerns in mind, we revisit our calculations to evaluate their suitability in the current quantum computing ecosystem and scalability for future developments and applications.

We first consider the circuit complexity of the ADAPT-VQE ansatz in our calculations in Fig.~\ref{fig:cnots}.
We show the number of \(T\) and (sequential) CNOT gates, where the later circuits are optimized to have gates run in parallel, versus the error on the exact energy for the iteratively improved ADAPT-VQE ansätze in our calculations of the deuteron and $^3$He.
For the deuteron, reaching 1~MeV precision requires an ansatz containing on the order $10^3$ (sequential) CNOT gates and \(T\) gates.
For $^3$He, the same precision requires about an order of magnitude more.
This increase is partially due to the more complicated operators on the $n_q =3L^3$ lattice, but also partially due to the increased number of exponentials \Nexp{} required to reach the same precision.
Going to $n_q=4L^3$ or to larger lattice extents $L$, we expect to see similar increases based on the increased basis size.
For a fixed basis size and increasing system size $A$, however, we expect this to scale more mildly as the operator pool (and the associated operator complexity) is fixed.
We see this also for $^3$He, where \Nexp{} is larger than for the deuteron for the same precision.
As our error decreases exponentially with increasing \Nexp{}, reaching higher precision generally does not require substantially more complex circuits.

Next, we consider the cost of repeated measurements on  hardware.
Based on the estimates in Sec.~\ref{sec:SampleComplexity} and Fig.~\ref{fig:shot_scaling} and our simulations including measurement noise, we find that reaching 1~MeV precision requires (at most) on the order of $10^4$--$10^6$ measurements per optimization step.
This cost scales mildly with system size and lattice size, but is already important.
Based on these findings, ADAPT-VQE appears more attractive as a tool for preparing approximate initial states, since the optimization need not reach unit fidelity with the ground state. Instead, once a sufficient overlap (e.g., 10\%) is obtained, the state can be passed to fault-tolerant quantum algorithms with guarantees, such as QPE or other filtering methods, which can then extract the ground-state component.
We find in Figs.~\ref{fig:h2_no_noise} and~\ref{fig:he3_no_noise} that our ADAPT-VQE ansätze effectively produce states with high fidelity with the ground state, hinting that our scheme could serve as a basis for preparing nuclear ground states on larger lattices. 
For example, for the deuteron (triton) we reach a fidelity greater than 0.6 for $\Nexp = 3$ ($\Nexp = 6$). We refer the reader to Ref.~\cite{roggero2020}, further improved in Ref.~\cite{spagnoli2025quantum}, where the cost of simulating the time evolution of a similar Hamiltonian has been estimated, both using Trotter-Suzuki product formula and qubitization.

Ultimately, the approach we pursue in this work has a few key features that make it attractive for future quantum computations.
Formulating things in second quantization means that the number of qubits required is relatively high, which restricts current applications to lattice extents $L=2, 3$.
Realistic lattice sizes of $L\sim 10$ will require several thousand qubits.
At the same time, however, the choice of lattice basis makes the Hamiltonian extremely sparse, exploiting the short-range nature of nuclear forces.
This makes the construction of physics-informed ansätze and the evaluation of expectation values and gradients relatively simple, achievable on hardware that is (or will soon become) available.
As this approach is refined and at the same time larger quantum computers become available, it will naturally scale and make more sophisticated computations possible, similar to the progress in classical computations of nuclei driven by scalable many-body methods and continuously increasing computational power.

\section{Summary}
\label{sec:Outro}

We developed quantum computations of nuclei on a coordinate-space lattice.
This formulation naturally exploits the short-range nature of nuclear forces, leading to very sparse Hamiltonian matrices.
This sparsity makes this approach well suited to quantum computations, as the number of Pauli terms required to evaluate Hamiltonian expectation values scales like the lattice volume $L^3$, which is proportional to the number of qubits $n_q$.

Using ADAPT-VQE, we performed simulations of quantum computations of the deuteron and $^3$He with a simple Hamiltonian from pionless effective field theory including two- and three-nucleon forces.
We found that our calculations are able to systematically converge toward the exact ground state results as we expand the complexity of the variational ansatz used in ADAPT-VQE.
Our calculations also performed well in simulations involving a finite number of stochastic measurements of the quantum circuit.

We presented scaling trends in lattice size $L$ and system size $A$ for the number of shots required and also investigate the circuit complexity (in terms of number of CNOT and \(T\) gates) for our calculations.
We found that the computational costs of our approach generally scale like the number of lattice sites and more mildly in the mass number $A$.
Based on this, combining the lattice calculations with ADAPT-VQE can be an efficient, scalable approach to computing high quality initial states with considerable overlap with the exact ground state.
Such initial states could then be used as the starting point for QPE calculations, where the fact that our initial state is relatively easy to construct and the Hamiltonian is very sparse will be essential advantages.

\section*{Data availability}

The data shown in the figures and table of this work are available in Ref.~\cite{gu_2025_17279920}.

\acknowledgments
This work was supported by the U.S. Department of Energy, Office of
Science, Office of Nuclear Physics, under Award No.~DE-FG02-96ER40963 and No.~DE-SC0021642, by the Quantum Science Center, a National Quantum Information Science Research Center of the U.S. Department of Energy, and by the Laboratory Directed Research and Development Program of Oak Ridge National Laboratory, managed by UT-Battelle, LLC, for the U.S.\ Department of Energy.  Oak Ridge National Laboratory is supported by the Office of Science of the Department of Energy under contract No.~DE-AC05-00OR22725. 
This research used resources of the Oak Ridge Leadership Computing Facility located at Oak Ridge National Laboratory. 
The authors gratefully acknowledge the Gauss Centre for Supercomputing e.V.\ (www.gauss-centre.eu) for funding this project by providing computing time through the John von Neumann Institute for Computing (NIC) on the GCS Supercomputer JUWELS at J\"ulich Supercomputing Centre (JSC). OK is supported by CERN through the CERN Quantum Technology Initiative.

This manuscript has been authored in part by UT-Battelle, LLC, under contract DE-AC05-00OR22725 with the US Department of Energy (DOE). The US government retains and the publisher, by accepting the article for publication, acknowledges that the US government retains a nonexclusive, paid-up, irrevocable, worldwide license to publish or reproduce the published form of this manuscript, or allow others to do so, for US government purposes. DOE will provide public access to these results of federally sponsored research in accordance with the DOE Public Access Plan (\url{http://energy.gov/downloads/doe-public-access-plan}).

\appendix
\section{Hamiltonian}
\label{app:ham}
The Hamiltonian used in this work depends on two coupling constants, one ($v$) for the on-site two-body contact and one ($w$) for the on-site three-body contact. The coupling constants were adjusted by performing exact diagonalizations. Because of the simplicity of our Hamiltonian (and the small sizes of lattices used in the quantum computations) we only aimed for a reproduction of some qualitative features of atomic nuclei, i.e., a weakly bound two-body system and a stronger bound three-nucleon system. The SU(4) symmetry in the two-body sector then also leads to a bound neutron-neutron (and proton-proton) system, and one cannot distinguish between $^3$H and $^3$He. Results are shown in Table~\ref{tab:systems}. The lattice spacing of $a=2$~fm corresponds to momentum cutoffs of about 0.3~GeV and is similar of that taken in nuclear lattice effective field theory~\cite{epelbaum2011}. We note that energies increase with increasing lattice extent $L$ (keeping all other parameters fixed) because there is less energy to be gained from tunneling to a periodic copy of the lattice. The corresponding  finite-size corrections of energies are well understood~\cite{luscher1985,konig2017} and not dealt with here.

\begin{table}
\renewcommand{\arraystretch}{1.2}
\centering
\caption{Ground-state energies of $^2$H and $^3$He (in MeV) from exact diagonalizations on lattices of extent $L$, lattice spacing $a=2$~fm, two-body coupling $v=-9.0$, and three-body coupling $w=10.8$. }
\label{tab:systems}
\begin{ruledtabular}
\begin{tabular}{dddd}
 L &  ^2{\rm H} & ^3{\rm He} \\ 
 \colrule
2 & -12.874 & -29.468   \\
3 &  -6.177 & -15.694   \\
4 &  -3.874 & -11.123   \\
5 &  -2.824 &  -9.561   \\
6 &  -2.265 &  -9.088   \\
\end{tabular}
\end{ruledtabular}
\end{table}

\section{ADAPT-VQE operator pool}
\label{app:operators}

For the pool of operators for our ADAPT-VQE calculations, we focus on a pool of operators producing orthogonal transformations.
We found that for our real-valued Hamiltonian this is more effective at generating appropriate transformations that lower the energy expectation value than general unitary transformations.
This can be understood through the following argument:

Assume we have a particle in a state 
\begin{equation}
    |j\rangle = \hat{a}^\dagger_j|\slashed{0}\rangle
\end{equation}
using second quantization and the vacuum state $|\slashed{0}\rangle$. 
One can now act with the operator $(j \ne k)$
\begin{equation}
\label{eq:A_type_operator}
    \hat{A}(\theta)\equiv 
    \exp{\left(\theta\left[\hat{a}^\dagger_k \hat{a}_j - \hat{a}^\dagger_j \hat{a}_k\right]\right)} \ ,
\end{equation}
or with 
\begin{equation}
    \hat{B}(\phi)\equiv 
    \exp{\left(i\phi\left[\hat{a}^\dagger_k \hat{a}_j + \hat{a}^\dagger_j \hat{a}_k\right]\right)}
    \label{eq:B_type_operator}
\end{equation}
onto the state $|j\rangle$. Here, $\theta$ and $\phi$ are both real  numbers. The operator~(\ref{eq:A_type_operator}) generates an orthogonal transformation (a Givens rotation), while the operator~(\ref{eq:B_type_operator}) generates a unitary transformation via the exponentiation of a Hermitian generator multiplied with a purely imaginary number. 
In the first case, we have
\begin{equation}
    \hat{A}(\theta)|j\rangle = \cos\theta |j\rangle +\sin\theta |k\rangle  \,,
\end{equation}
while in the second case, we have
\begin{equation}
    \hat{B}(\phi)|j\rangle =  \cos\phi |j\rangle +i\sin\phi |k\rangle\,.
\end{equation}
The resulting energy expectation values are
\begin{align}
\label{Atheta}
    E(\theta) &\equiv \langle j|\hat{A}^\dagger(\theta) \hat{H} \hat{A}(\theta)|j\rangle \nonumber\\
&=  \langle j|\hat{H}|j\rangle \cos^2\theta + \langle k|\hat{H}|k\rangle \sin^2\theta \nonumber\\
&+\left(\langle j|\hat{H}|k\rangle+\langle k|\hat{H}|j\rangle\right) \sin\theta\cos\theta  \ , 
\end{align}
and 
\begin{align}
\label{Bphi}
    E(\phi) &\equiv \langle j|\hat{B}^\dagger(\phi) \hat{H} \hat{B}(\phi)|j\rangle \nonumber\\
    &= \langle j|\hat{H}|j\rangle \cos^2\phi + \langle k|\hat{H}|k\rangle \sin^2\phi  \nonumber\\
&+i \left(\langle j|\hat{H}|k\rangle-\langle k|\hat{H}|j\rangle\right) \sin\phi\cos\phi \,.
\end{align}
For real-symmetric Hamiltonians $\langle j|\hat{H}|k\rangle=\langle k|\hat{H}|j\rangle$ and the last line in Eq.~(\ref{Bphi}) vanishes. Thus the energy will satisfy $E(\phi)< E(0)$ only if $\langle k|\hat{H}|k\rangle < \langle j|\hat{H}|j\rangle$. In contrast, the operator $\hat{A}(\theta)$ lowers the energy when acting on the state $|j\rangle$ for any $\langle k|\hat{H}|j\rangle\ne 0$. To see this, one can take $|\theta|\ll 1$ in Eq.~(\ref{Atheta}) and choose its sign such that $\theta \langle k|\hat{H}|j\rangle < 0$.
We clearly see that in this case it is generally easier to reduce the energy of our initial state through the application of an appropriately parametrized orthogonal transformation.

For our operator pool, we construct operators as in Eq.~\eqref{eq:A_type_operator}.
We include both one- and two-body operators.
Our one-body operators have the same structure as the kinetic energy, allowing a single nucleon to hop from one lattice site to a neighboring site via
\begin{equation}
    \hat{A}(\theta) = \exp{\left(\theta \hat{a}_{\mathbf{l}\tau s}^\dagger \hat{a}_{\mathbf{l}'\tau s} - \mathrm{H.c.}\right)} \ , 
\end{equation}
with nearest neighbors $\langle \mathbf{l}, \mathbf{l}' \rangle$.
For our two-body operators, we include all operators that move two nucleons with given spin and isospin projections to neighboring lattice sites:
\begin{equation}
\hat{A}(\theta) = \exp\left( \theta \, \hat{a}^\dagger_{\mathbf{l} \tau_1 s_1} \hat{a}^\dagger_{\mathbf{k} \tau_2 s_2} \hat{a}_{\mathbf{k}' \tau_2 s_2} \hat{a}_{\mathbf{l}' \tau_1 s_1} - \text{H.c.} \right) \ .
\end{equation}
Here $\mathbf{l}'$ and $\mathbf{k}'$ denote the original lattice sites, and $\mathbf{l}$ and $\mathbf{k}$ are the target lattice sites after the transition.
We always consider operators with $\mathbf{k}'=\mathbf{l}'$.
These operators are categorized into three types based on the movement pattern. Type I operators describe correlated pair-hopping in which both nucleons move together to the same adjacent site ($\mathbf{k} =\mathbf{l}$). Type II operators describe that one nucleon remains stationary while the other moves to an adjacent site ($\mathbf{k}=\mathbf{k}'$ or $\mathbf{l}=\mathbf{l}'$). Type III operators describe that both nucleons move separately to different neighboring sites ($\mathbf{k} \neq \mathbf{l}$).

For the deuteron, we only use the operators from Type I and Type II, and our deuteron operator pool allows for hopping beyond nearest-neighbors: transitions between any two distinct sites are allowed. Therefore we have in total 24 one-body operators, 28 Type I operators and 112 Type II operators in our pool.

For $^3$He, unlike the deuteron case, we include all three types of two-body operators to account for more complex interactions. We use the full two-body operator pool for the $^3$He system, allowing hopping between all lattice sites rather than restricting to nearest neighbors. This results in a total of 36 one-body operators and, specifically, 84 Type I operators, 168 Type II operators, and 336 Type III operators. For comparison, if we  restrict the two-body hopping to nearest-neighbor sites, the numbers are significantly smaller: 36 Type I, 72 Type II, and 36 Type III operators.
We performed such a simulation for $^3$He and found similar convergence behavior and the same final result as in Fig.~\ref{fig:he3_no_noise}. This indicates that both pools are sufficient for our ADAPT-VQE calculations and the adaptive operator selection is effectively able to identify the most essential operators to build an efficient variational ansatz.

\section{Translationally invariant initial state}
\label{app:trans}

The simple way to define an initial state on the lattice,
placing nucleons on definite lattice sites,
is not translationally invariant.
We outline how one could generally construct a translationally invariant initial state given some simple product state.

Let us consider the deuteron. For the initial state we occupy a single lattice site with a proton and a neutron, i.e., the state is 
\begin{equation}
|0,1\rangle \equiv a^\dagger_{0} a^\dagger_{1}|\slashed{0}\rangle \,.
\end{equation}
Here, we simply neglected spin degrees of freedom and essentially have states with even numbers for neutrons and states with odd numbers for protons. We obtain a translationally invariant state as follows. We introduce the anti-Hermitian deuteron hopping operator
\begin{equation}
    A_{l\to k} \equiv a^\dagger_{2k} a^\dagger_{2k+1}a_{2l+1} a_{2l} - a^\dagger_{2l}  a^\dagger_{2l+1} a_{2k+1} a_{2k}
\end{equation}
that moves the deuteron from lattice site $l$ to site $k$. 
The state 
\begin{equation}
\label{initial}
    |\psi_0\rangle = \exp{\left(\theta_{n-1} A_{0\to n-1} \right)}\cdots \exp{\left(\theta_1 A_{0\to 1} \right)} |0,1\rangle
\end{equation}
is then translationally invariant on a lattice with $n$ sites. 
Here, the angles have to be chosen such that 
\begin{equation}
\begin{aligned}
\label{angles}
    \sin\theta_k &= \frac{1}{\sqrt{n-k+1}} \ , \\
    \cos\theta_k &= \sqrt{\frac{n-k}{n-k+1}} \ .
\end{aligned}
\end{equation}
To see that this is correct, we note that 
\begin{equation}
    \begin{aligned}
    \exp{\left(\theta_1 A_{0\to 1} \right)} |0,1\rangle = \cos\theta_1 |0,1\rangle +\sin\theta_1 |2,3\rangle \ .
    \end{aligned}
\end{equation}
Repeated application of this operation, as demanded by Eq.~(\ref{initial}), then yields
\begin{align}
        |\psi_0\rangle &= \left(\prod_{l=1}^{n-1}\cos\theta_l\right) |0,1\rangle \nonumber\\
        &+  \sum_{k=1}^{n-1}\sin\theta_k \left(\prod_{l=1}^{k-1}\cos\theta_l\right) |2k,2k+1\rangle \nonumber\\
        &=  n^{-1/2}\sum_{k=0}^{n-1} |2k,2k+1\rangle \,.
    \end{align}
In the last step, we employed the angles~(\ref{angles}). Clearly, the state $|\psi_0\rangle$ is a coherent superposition of states where a deuteron is on each lattice site. It is also a correlated state. The effort of its preparation is ${O}(n)$. In practice, however, the preparation can only be implemented approximately, because  Eq.~(\ref{initial}) requires a Trotterization. As is seems questionable to implement the exact symmetry corresponding to translational invariance only approximately, we did not pursue this further.

\bibliography{qc}

\end{document}